# Investigation of light ion fusion reactions with plasma discharges

T. Schenkel*[,1], A. Persaud[1], H. Wang[1], P. A. Seidl[1], R. MacFadyen[1], C. Nelson[1], W. L. Waldron[1], J.-L. Vay[1], G. Deblonde[2], B. Wen[3], Y.-M. Chiang[3], B. P. MacLeod[4], and Q. Ji[1]

[1]Accelerator Technology and Applied Physics Division, Lawrence Berkeley National Laboratory, Berkeley, CA 94720, USA

[2]Chemical Sciences Division, Lawrence Berkeley National Laboratory, Berkeley, CA 94720, USA

[3]Department of Materials Science and Engineering, Massachusetts Institute of Technology, Cambridge, MA 02139, USA

[4]Department of Chemistry, University of British Columbia, Vancouver, British Columbia, Canada

*corresponding author, t_schenkel@lbl.gov

**Abstract**

The scaling of reaction yields in light ion fusion to low reaction energies is important for our understanding of stellar fuel chains and the development of future energy technologies. Experiments become progressively more challenging at lower reaction energies due to the exponential drop of fusion cross sections below the Coulomb barrier. We report on experiments where deuterium-deuterium (D-D) fusion reactions are studied in a pulsed plasma in the glow discharge regime using a benchtop apparatus. We model plasma conditions using particle-in-cell codes. Advantages of this approach are relatively high peak ion currents and current densities (0.1 to several A/cm$^2$) that can be applied to metal wire cathodes for several days. We detect neutrons from D-D reactions with scintillator-based detectors. For palladium targets, we find neutron yields as a function of cathode voltage that are over 100 times higher than yields expected for bare nuclei fusion at ion energies below 2 keV (center of mass frame). A possible explanation is a correction to the ion energy due to an electron screening potential of 1000±250 eV, which increases the probability for tunneling through the repulsive Coulomb barrier. Our compact, robust setup enables parametric studies of this effect at relatively low reaction energies.

## 1. Introduction

Studies of light ion fusion are important for our understanding of stellar fuel chains and for the development of fusion technologies [1]. Ion beams and cold targets have long been used to determine fusion reaction cross sections and yields [1]. Controlled experiments with hot plasma targets are now within reach for the first time [2, 3]. For energies well below the Coulomb barrier, $E_c$, ($E_c > 400$ keV for fusion reactions between hydrogen isotopes), the reaction cross section, $\sigma(E)$, decreases exponentially with decreasing ion kinetic energy, $E$, and is often expressed as

$$\sigma(E) = S(E) \cdot \frac{1}{E} \cdot exp\left(-\frac{B_G}{\sqrt{E}}\right)$$
(Eq. 1)

the product of the astrophysical S-factor, S(E), a geometric factor, 1/E, and a screening factor that expresses the exponential energy dependence of the Coulomb barrier penetrability, with Gamow factor, $B_G= \pi \alpha Z_1 Z_2 (2 m_r c^2)^{\frac{1}{2}}$, speed of light, c, atomic numbers of the reacting nuclei, $Z_1$, $Z_2$, reduced mass $m_r= m_1 m_2/(m_1+m_2)$, and the fine-structure-constant $\alpha=1/137$ [4]). For the D-D reaction the center of mass energy, $E_{cm}$, is one half of the kinetic energy of deuterium ions in the laboratory frame, $E_{lab}$, that strike a stationary target.

Screening effects in gases, solids, and dense plasmas can increase fusion rates at low reaction energies by several orders of magnitude because screening of the repulsive Coulomb potential by plasmas or target atom electrons increases the probability for ions to tunnel through the Coulomb barrier [1]. The electron screening effect can be expressed as a screening potential, $U_e$, that is an effective correction to the bare-nucleus reaction energy; with ion kinetic energy, $E_k$, the modified reaction energy then becomes: $E=E_k+U_e$. The electron screening effect is negligible for ion energies near or above the Coulomb barrier; but, due to the exponential dependence of barrier penetrability on ion energy, screening effects can increase fusion reaction rates at relatively low reaction energies by several orders of magnitude. This phenomenon depends on the electron or plasma density and is present in many stars [1].

A series of experimental studies have been designed to quantify electron screening effects in gases and solid targets. Reported values of screening potentials in the D-D reaction for a series of target materials range from under 30 eV to over 800 eV [6-9], the latter being much larger than the value for gas targets of 27 eV [10]. Experiments with hot, dense plasmas that reproduce stellar conditions in the laboratory have come within reach in recent years [2, 3]. But these still require large facilities with limited access and relatively low shot rates. Metal hydrides can be viewed as simple analogs of dense plasmas [11]. While experiments that reproduce stellar conditions in the laboratory are preferable, experiments with metal hydride targets as analogs of dense plasmas enable access to relevant aspects of the physics of low energy fusion reactions in low-cost, bench-top experiments. Resonances in nuclear reaction cross sections at low reaction energies can also lead to reaction rate changes that are missed when relying on extrapolations of astrophysical S factors based on high energy data alone (reference [12] is a recent example).

The standard approach to measuring nuclear reaction cross sections is to have a beam of ions (with well-defined ion species, current and kinetic energy) impinge on a sample that contains the target nuclei of known density. The sample can be pre-loaded or beam loaded with the target nuclei. Ion beam approaches, directly or through variants such as the Trojan Horse method [12, 13], have been successful in extending our knowledge of nuclear cross sections into the Gamow window of stellar systems [14]. But the ion currents available have been mostly limited to below 1 mA, especially for low energy experiments. Facilities with higher beam intensities have been proposed or are under development [14]. As an alternative to well defined ion beams, energetic ions from Z-pinches, plasma Hall accelerators [15] and plasma discharges [8] have also been used to study light ion fusion. Lipson *et al.* [8] reported on D-D fusion studies with currents and current densities of ~0.5 A and 0.5 A/cm$^2$, respectively, in a parallel plate geometry with a glow discharge plasma established between the plates. The energy range was extended to as low as $E_{cm}$=0.4 keV, and an electron screening potential of $U_e$=610±150 eV was reported from measurements of D-D fusion rates with detection of protons.

In this article, we report results from plasma discharge experiments with a metal wire cathode in the glow discharge regime [16]. This technique enables the measurement of light ion fusion yields and reaction branching ratios in a compact, economical apparatus. This allows parametric studies of factors that can affect reaction rates, such as electron screening potential, ion dose rate, target loading conditions and the presence of (transient) defects in metal hydride lattices.

## 2. Experimental setup

A schematic of our setup is shown in Figure 1. We use a standard stainless steel cube with an edge length of 152 mm as the plasma chamber. The base pressure is in the mid $10^{-7}$ Torr range. A 5 cm long metal wire (*e.g.*, palladium or titanium) connected to the negative terminal of a high voltage pulser acts as the cathode. Wire diameters are 0.5 and 1 mm. The wire is surrounded by a stainless steel cage which acts as the anode, and is grounded through a current transformer. Cage diameters are 1.25 and 2.5 cm. The high voltage pulser is a charged capacitor and IGBT (insulated gate bipolar transistor) array that delivers 1 to 5 kV square-wave pulses to the wire at a repetition rate of up to 50 Hz with 20 μs pulses, for a duty cycle of $10^{-3}$. For higher discharge biases, we added a step-up transformer to the pulser circuit. We operate the plasma in the glow discharge regime with deuterium gas ($D_2$) pressures in the range of 0.1 to 2 Torr. Control experiments and background runs were conducted with regular hydrogen gas ($H_2$). For lower deuterium gas pressures, we found it difficult to strike a plasma; for higher deuterium pressures, the glow discharge became unstable and developed into an arc discharge. During the discharge, positive deuterium ions were accelerated across the plasma sheath into the wire cathode. Deuterium ions were implanted into the wire; consecutive ions can undergo fusion reactions.

We measured light emission from the plasma with an Ocean Optics Flame Series fiber-coupled spectrometer and confirmed the presence of deuterium discharge by observing the Balmer line series (400 to 660 nm). Optical emission spectroscopy is useful as a basic plasma diagnostic, and to track potentially present light emission from metal ions and excited atoms, which can indicate arcs or excessive heating of the cathode wire.

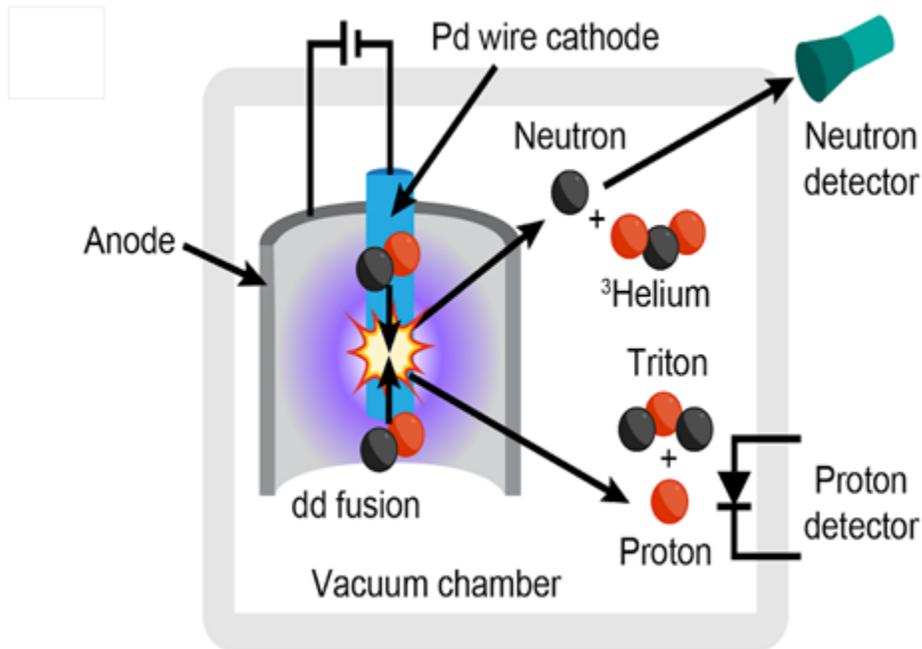

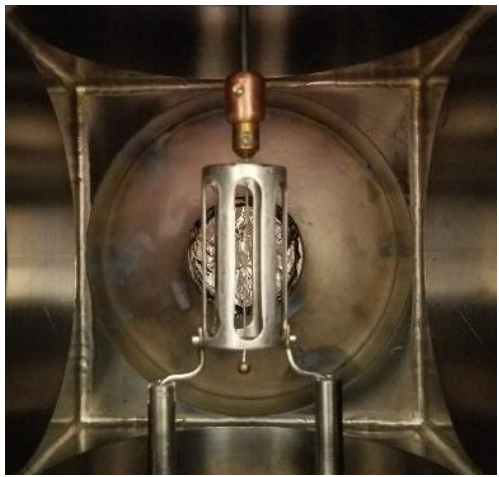 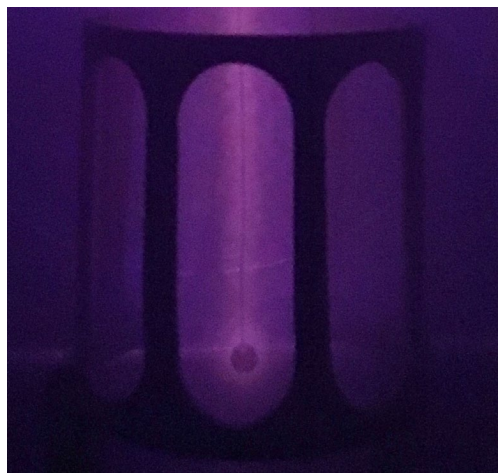

**Figure 1**: Schematic, top, and photo, bottom left, of the glow discharge setup in a stainless steel cube (15.2 cm edge length) with palladium wire cathode (0.5 mm diameter) in a stainless steel cage anode (2.5 cm diameter). Photo of light emission from a deuterium plasma, bottom right.

Typical voltage and current traces during a pulse are shown in Figure 2. Once tuned, discharges were stable for days, enabling extended runs where over 10 Coulombs of ions impinge on a wire target for fluences in the $10^{21}$ D/cm² range.

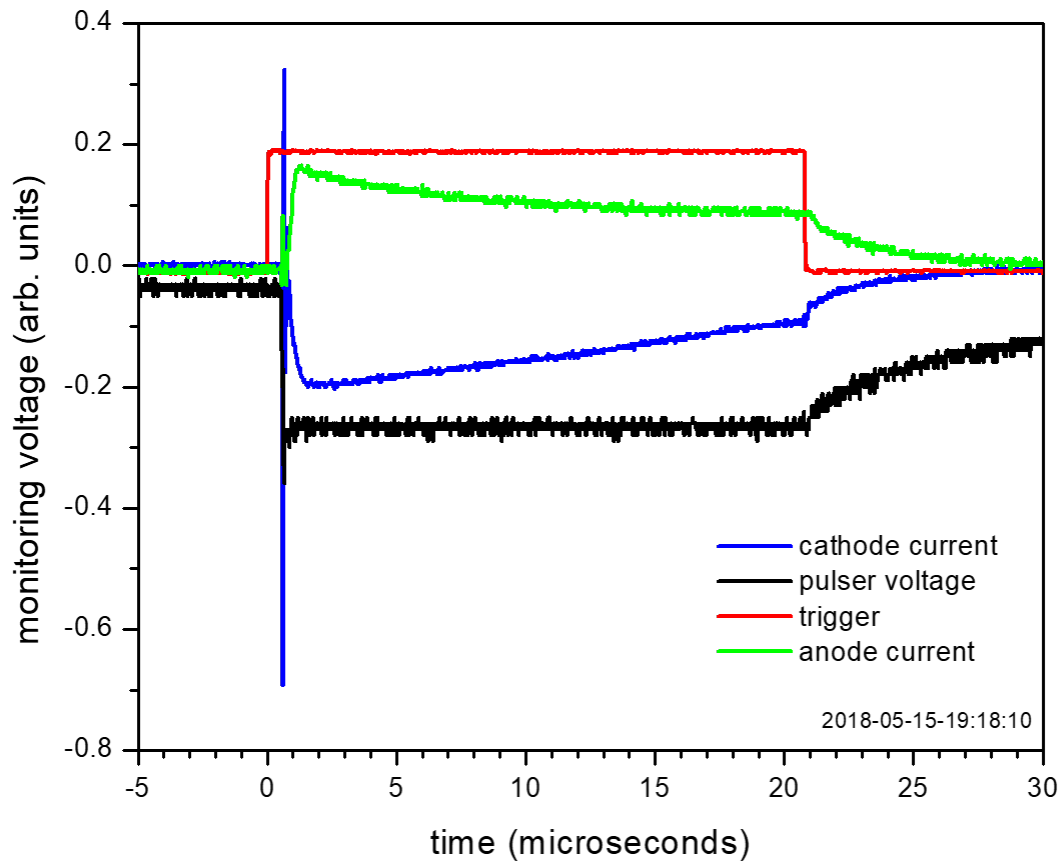

**Figure 2**: Traces of monitoring voltages that track the plasma discharge pulser voltage (black), anode cage current (green) and cathode wire current (blue) at a cathode voltage of 2 kV, deuterium gas pressure of 2 Torr, and palladium wire diameter of 0.5 mm. In this example, the cathode current (blue) varied between 1 and 2 A during the discharge.

In Figure 3, we show a typical current-voltage curve of the plasma discharge for a deuterium gas pressure of 0.5 Torr with a palladium wire cathode. We tuned the discharge bias, current and pressure conditions to achieve stable glow discharges for extended (≫1 h) operation.

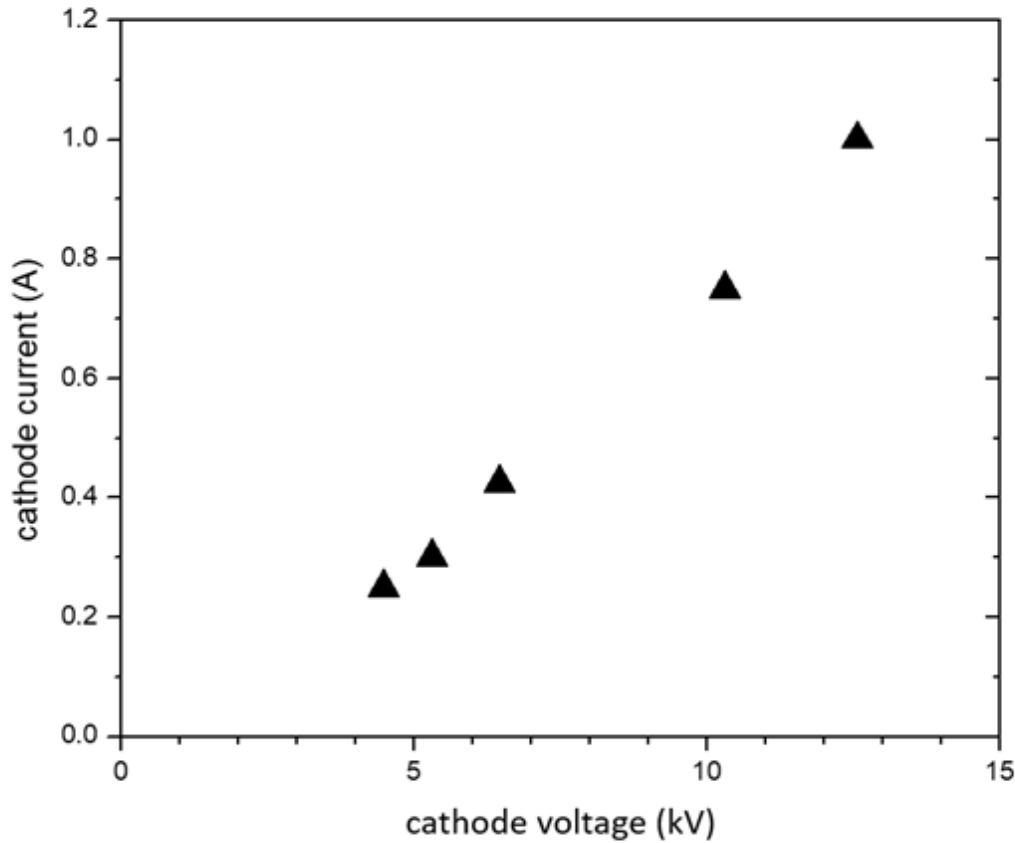

**Figure 3**: Typical current-voltage curve of the plasma discharge in the glow discharge regime with a palladium wire (0.5 mm diameter) at a deuterium gas pressure of 0.5 Torr.

From Monte Carlo simulations (SRIM [17]), we estimate the range of deuterium ions in palladium to be from 10 to 100 nm for deuterium ion energies from 2 to 12 keV. Energy dependent backscatter yields are 10 to 25% of incident ions. For a given ion current density and duty cycle, incident deuterium ions can load into the palladium lattice. However, we have not yet implemented methods to track target loading *in situ*. Rates of diffusion of deuterium into the bulk of the wire and emission into the vacuum are unknown for the conditions of pulsed deuterium ion flux and relatively low deuterium gas pressures. These uncertainties in target loading conditions make determination of absolute nuclear reaction yields and extraction of absolute reaction cross sections very challenging. By comparing neutron detection rates for a series of discharge biases, we measure how the reaction yields scale, and then compare this to the scaling of reaction rates from experimental data in the literature [5-10, 18] and to predictions from theory [4]. Changes in the deuterium ion distribution with changing plasma discharge conditions also have to be considered. With the constraints currently present, relative yields and trends can be measured.

## 3. Simulations of ion species and energy distributions

We did not directly measure ion species and ion energy distributions that impinged on the cathode wires during our plasma discharge experiments. Lipson *et al.* [8] argue that most ions have the full energy corresponding to the applied discharge bias, based on the low degree of ionization in the plasma and the assumption of low collision rates during acceleration of ions across the plasma sheath for glow discharges with pressures ~1 Torr. To address this for our experimental conditions, we ran particle-in-cell (PIC) simulations of the plasma with the WARP [19] and VSim codes [20], constrained by cross sections for the dominant elastic and inelastic collision processes (Table 1). The energy distribution of ions arriving at the cathode is determined by the acceleration of ions caused by the potential drop from where the ions are formed (usually in the bulk plasma, where the plasma potential is $V_p \approx 10$ V) across the plasma sheath to the cathode at -12.4 kV < $V_c$ < -2.4 kV. Since the bulk plasma is quasi-neutral, most of the potential drop occurs in the sheath. The sheath thickness is determined by the plasma temperature, density, voltage bias, and collisional effects. For our experimental conditions, we find a sheath thickness of <1 mm from the simulations, a small fraction of the anode-cathode distance for a cage diameter of 1.25 or 2.5 cm. The expected maximum ion kinetic energy, $E_{max}$, is given by the applied cathode bias, $V_c$, with a small correction from the bulk plasma potential, $V_p$. $E_{max} \approx q_e \cdot \Delta V = q_e (V_p - V_c)$. Here $q_e$ is the elementary electric charge. Broadening of this ion energy due to the plasma ion temperature's Maxwellian distribution is negligible compared to the total ion energy. But in the ~1 Torr pressure regime, there is a significant likelihood that the ions will interact with the background neutral gas atoms and molecules, leading to momentum and energy transfer as well as charge exchange reactions which reduce the ion kinetic energy below this $E_{max}$ value.

To estimate the energy distribution of the ions, and to quantify the effect of collisions and reactions of hydrogen ions, atoms and molecules, we have built a fully kinetic PIC model into the plasma simulations. But including all the possibly relevant reactions and collisions into PIC simulations of the plasma and background gas is computationally expensive. Hence, we initialized our plasma simulation using a "global" plasma model approach [21]. We used data for hydrogen in our simulations; the reactions listed in Table 1 are included in our model, together with cross-sections from [22] and [23].

The result of the global model analysis is shown in Figure 4 for a hydrogen gas pressure of 0.5 Torr. The density of electrons and protons increases with increasing discharge power (*i.e.*, the product of discharge bias and cathode current). When the discharge power increases from 1.1 to 12.6 kW, the electron temperature increases slightly from 2.5 to 2.9 eV. We find that protons are always the most abundant ion species in these simulated glow discharge plasmas. But at the lowest discharge powers, $H_3^+$ ions are also abundant at about 1/3 the number of protons. This implies that, for relatively low voltage discharges, fusion reactions of $D^+$ occur in the presence of a nearly equal flux of $D_3^+$ ions. The energy per nucleon in molecular ions is reduced corresponding to their higher mass, and the corresponding fusion reaction cross section is exponentially smaller. However, the impact of low velocity molecular ions affects the deuterium density and metal lattice defect structure, which could affect electron screening potentials and hence fusion reaction rates.

| | |
|---|---|
| e + $H_2$ → H + H + e [24] | $H^-$ + $H_2^+$ → $^3H$ [24] |
| e + $H_2$ → 2e + H + $H^+$ [25] | H- + $H_3^+$ → $^4H$ [24] |
| e + $H_2$ → 2e + $H_2^+$ [24] | e + $H^-$ → H + 2e [24] |
| e + $H_2$(v) → 2e + $H_2^+$(v=1-4) [26, 27] | e + $H_3^+$ → H + H + $H^+$ + e [24] |
| e + $H_2$ → H + H- [24] | e + $H_2^+$ → 2H [24] |
| e + $H_2$(v) → H + $H^-$(v=1-9) [28] | e + $H_3^+$ → 3H [24] |
| e + H → $H^+$ + 2e [24] | e + $H_3^+$ → H + $H_2$ [24] |
| H + $H_2^+$ → $H^+$ + $H_2$ [24] | e + $H_2$ → e + $H_2$ [29] |
| e + $H_2^+$ → e + $H^+$ + H [24] | e + $H_2$ → e + $H_2$(v) (v=1-6) [30, 31] |
| $H_2$ + $H_2^+$ → $H_3^+$ + H [24] | e + $H_2$(v) → e + $H_2$(v+1) (v=1-8) [32] |
| e + $H_3^+$ → $H_2^+$ + $H^-$ [24] | e + H → e + H [33] |
| $H^-$ + $H^+$ → H + H [24] | e + H → e + H(v)(v=1-5) [33] |
| | e + H→$H^-$ [24] |

Table 1: Reactions included in the plasma simulations (e = electrons).

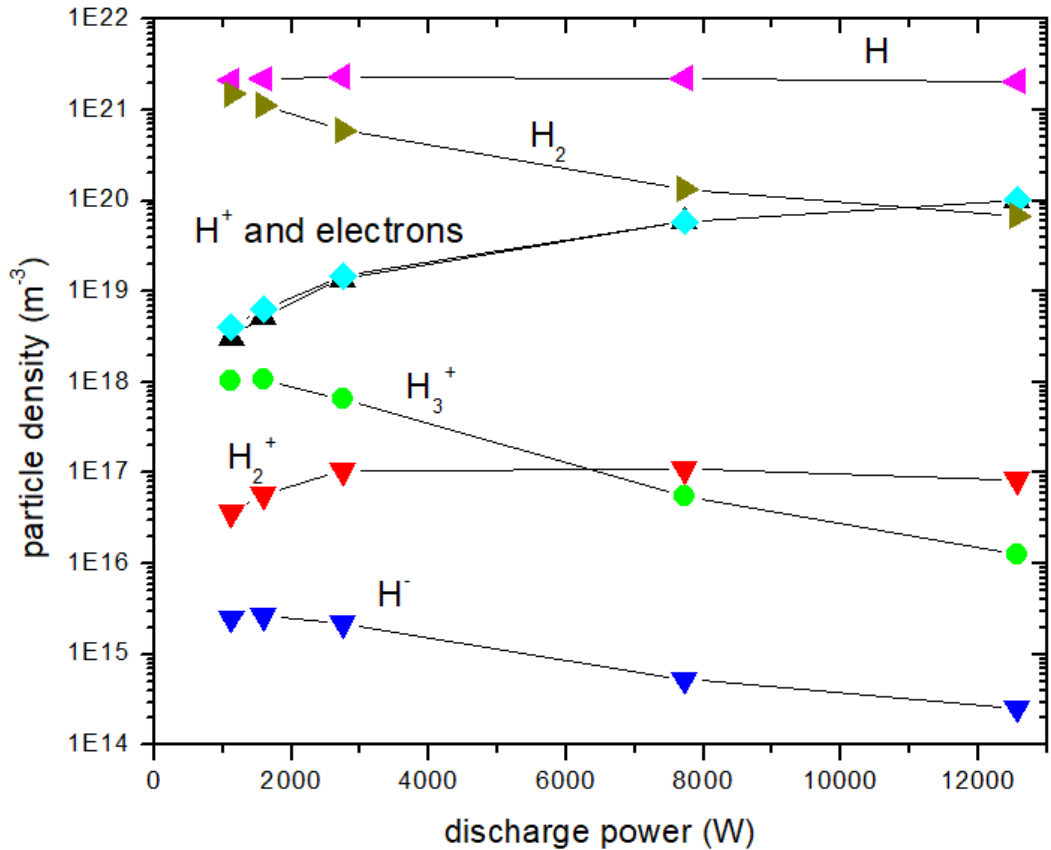

**Figure 4**: Simulated density of the main plasma species as a function of plasma power in the global model for a hydrogen gas pressure of 0.5 Torr. Trends (shown with straight lines between data points to guide the eye) were similar for a hydrogen gas pressure of 2 Torr.

Using this approach, we were able to increase the speed of the VSim PIC simulations [20] by initializing them with densities that are likely closer to equilibrium, and to limit the tracked collisions to the most significant ones; for example, charge exchange and momentum exchange in the sheath. We then applied this to simulations of the ion dynamics in the small-scale sheath (excluding most of the bulk plasma to save computation time) to estimate the ion energy distribution at the cathode. The resulting ion energy distributions, f(E), are shown in Figure 5 for the maximum and minimum cathode voltages that were used in our experiments with palladium wire cathodes.

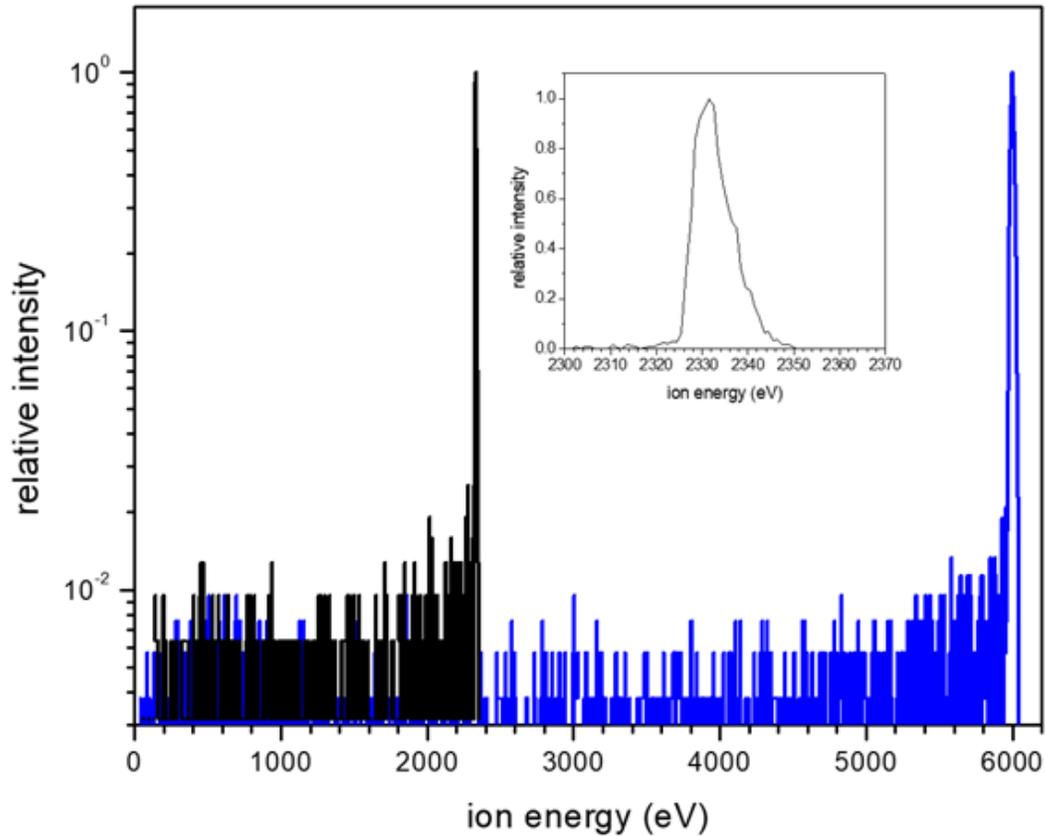

Figure 5: Simulated ion energy distribution f(E) for cathode voltage and hydrogen gas pressure of (a) -2.4 kV and 2 Torr, respectively and (b) -6.5 kV and 0.5 Torr, respectively. The insert shows the details of the peak of the simulated ion energy distributions for -2.4 kV.

In Figure 6, we show the resulting shifts in the ion energies as a function of cathode voltage. We show shifts in the peak ion energy, i. e. the maximum ion energy, $E_{max}$, and in the mean ion energy, from summing over the ion energy distribution and normalizing to the number of ions. We find that collisions in the sheath reduce the mean ion energy by $dE_{mean}$= 1.2 keV (or about 10 to 25% of the applied cathode voltage) at a discharge pressure of 0.5 Torr, while the peak ion energy, $E_{max}$, is reduced by $dE_{peak}$=0.5 keV. The ion energy distribution at the peak ion energies is broadened by 10 to 20 eV due to collisions in the sheath. This broadening is relatively small, about 10 to 20 eV, much smaller than the peak ion energies, $E_{max}$, of 2.3 and 12.1 keV. The simulation results for a series of cathode voltages in Figure 6 show significant scatter which reflects the limited accuracy of the simulations we performed.

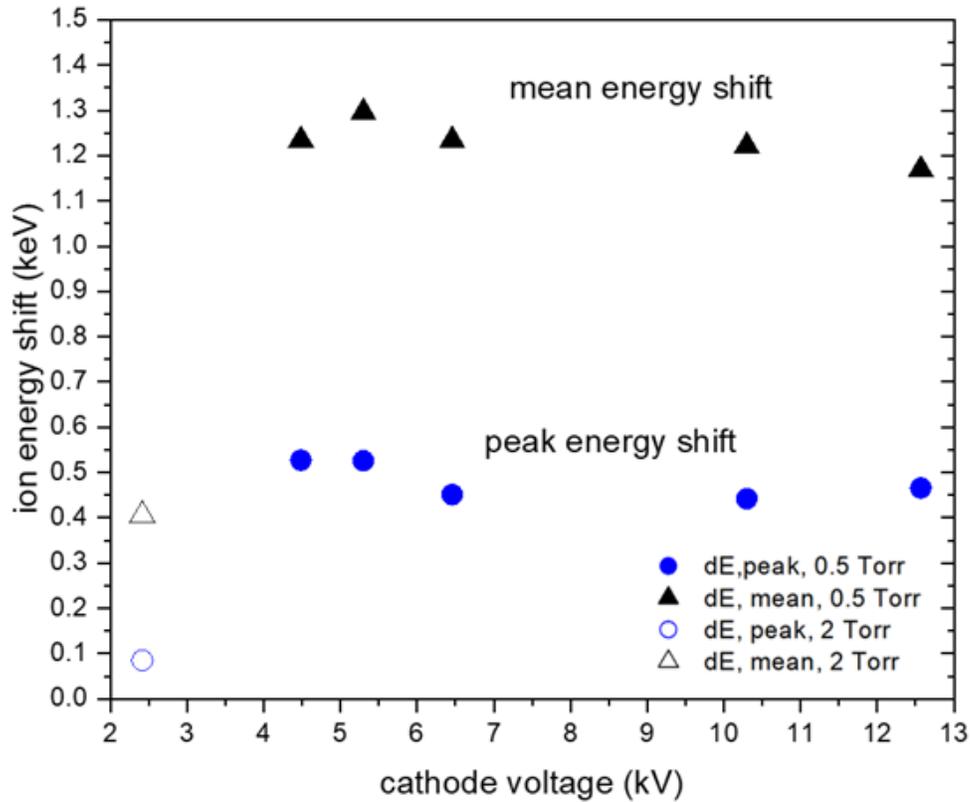

Figure 6: Results of simulations of the reduction of the peak and mean ion energies, $dE_{peak}$ and $dE_{mean}$, due to collisions in the plasma sheath as a function of cathode voltage for plasma discharges with a hydrogen pressure of 0.5 Torr (closed symbols) and two simulation data points for 2 Torr and a cathode voltage of 2.4 kV (open symbols)

The interplay of discharge power (cathode voltage and current) and hydrogen gas pressure leads to changes in the plasma sheath thickness. We simulated one example with a cathode voltage of 2.4 kV and discharge current of 4 A at 2 Torr and find a reduction in the sheath thickness that led to reduced ion energy shifts (open symbols in Figure 6).

For exponentially decreasing fusion cross sections as a function of ion energy, the population of atomic ions with the highest energies will dominate the observed fusion yields. In the plasma experiments here, the ion energy is determined by the cathode voltage, $V_c$, a small correction due to the plasma potential, $V_p$, and a correction due to collisions in the sheath. For comparison of experimental results with theoretical predictions on D-D fusion yield scaling, we use the ion energy distributions from our simulations to calculate expected thick target yields for bare nuclei and with a series of electron screening potentials, $U_e$. Direct measurements of the ion species and energy distribution would be preferable to support quantitative conclusions on fusion yield scaling and electron screening potentials.

Page 11 of 22

Compared to ion beam experiments (with well-defined ion energies and ion species), plasma discharge experiments enable probing the effects of much higher ion dose rates (ions/cm$^2$/s) on fusion reaction rates. Future improvements include refined plasma simulations with validation from benchmarking [34] as well as implementation of plasma diagnostics.

## 4. Detectors

We used two widely deployed types of neutron detectors to track D-D fusion rates during extended plasma discharge runs: $^3$He-based proportional counters, and scintillators coupled to a photomultiplier tube (PMT) [35]. The $^3$He-based detectors (Health Physics Instruments 6060) detect neutrons with much higher efficiency than gamma rays. Natural neutron and gamma ray background was attenuated by a shielding enclosure that included a layer of borated polyester and a sheet of lead. The $^3$He-based detector was useful for experiments with cathode voltages above 6 kV, but the detection efficiency and background rate precluded us from using it for lower cathode voltages. The fact that $^3$He-based detectors are insensitive to gamma rays makes them useful to support the analysis of data from detectors for which discrimination of neutron and gamma ray signals requires careful pulse shape discrimination.

The second neutron detector we implemented was based on a volume of liquid organic scintillator (Eljen Technology EJ-309) [36], where incident neutrons generate scintillation light through a series of scattering events. Light is detected in a photomultiplier tube (PMT). A photo of this detector next to the plasma chamber is shown in Figure 7.

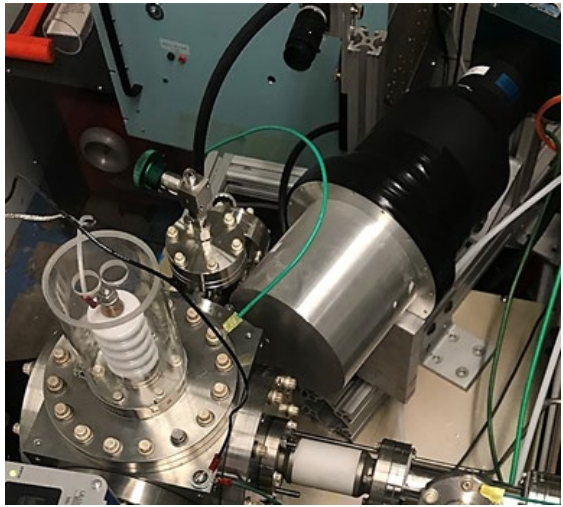

**Figure 7**: Photo of the EJ-309 based neutron detector with photomultiplier next to the plasma chamber.

The organic scintillator cartridge in the EJ-309-based detector assembly has a diameter of 15 cm and is placed 21 cm away from the wire cathode. The detection efficiency for 2.5 MeV neutrons is about 10% [36]. Examples of neutron signals detected over the course of extended (≫1 h) runs are shown in Figure 8 for a series of cathode voltages at a deuterium gas pressure of 0.5 Torr.

Compared to the $^3$He based detector, the EJ-309-based detector is more sensitive to neutrons but also to gamma rays. Pulse shape discrimination (PSD) is required to differentiate the two and for a low ratio of neutrons/gamma events the analysis protocol can introduce biases in the results [37, 38].

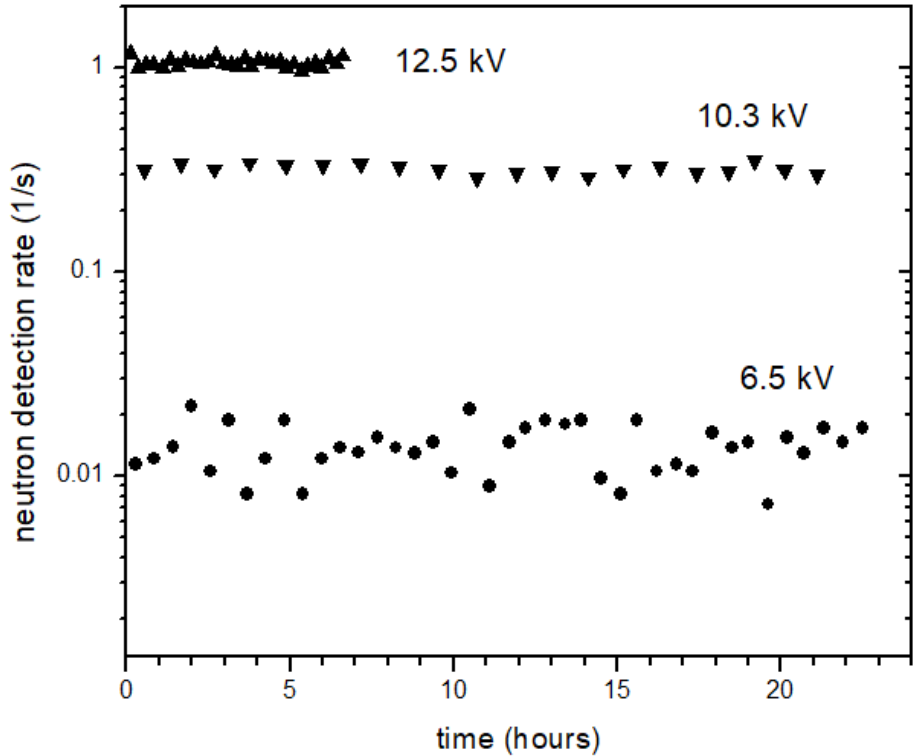

**Figure 8:** Neutron rates detected in the EJ-309-based detector as a function of time during extended plasma discharge runs with cathode voltages of 6.5 (circles), 10.3 (down triangles), and 12.5 kV (upward triangles). Glow discharge plasma conditions were stable leading to constant neutron production rates for continuous runs lasting for a few hours to several days.

We adopted two methods for PSD between neutrons and gamma rays for the analysis of fusion reaction yields in the neutron channel: charge-integration PSD and machine learning PSD. Charge integration PSD has been widely used for neutron/gamma ray discrimination, including with EJ-309 scintillator material [35-38]. Here, the full PMT signal is integrated and compared to the integrated charge in the tail of the pulse height distribution. The tail is defined empirically to commence 14 ns after the peak of the PMT pulse waveform.

In machine learning PSD, we adapted a label spreading algorithm. This semi-supervised training method was chosen because unlabeled data are easier to obtain than labeled data [39]. First, clearly distinguished pulses are labeled as neutron or gamma ray events based on a preliminary charge integration analysis as described above. Next, 20 components are extracted from the raw data via principal-component analysis (PCA). Finally, the components' values and initially labeled events are used as input to the label-spreading algorithm, to label initially unidentified pulses. Raw neutron data and the scripts we used for neutron data analysis will be made available upon request.

With neutron detectors we can track trends in the D+D→$^3$He (0.85 MeV) +n (2.45 MeV) branch of the D-D fusion reaction. In Figure 1, we also indicate the presence of a proton detector. Proton detectors have been widely used in earlier studies of the D-D reaction [5-10]. We have not yet succeeded in implementing proton detection to track 3 MeV protons from the D+D → H (3 MeV) + T (1 MeV) branch. Our first approach, based on silicon diode detectors, failed due to excessive electrical noise and induced charge signals that overwhelmed the detectors during plasma discharge pulses. We are now implementing a proton detector based on a scintillator coupled with a light guide to a PMT.

With implementation of both neutron and proton detectors, both dominant branches (i.e. the $^3$He+n and T+p) of D-D fusion can be tracked. Measurements of the branching ratio of light ion fusion reactions at low energies can shed light on hypothetical threshold resonances and reaction channels that have to date not been quantified at very low reaction energies, below $E_{cm}$ = 3 keV [16, 40, 41]. Future studies can also include gamma ray detectors to probe any potential changes in the relative contribution of the usually very weak $^4$He + gamma ray branch of the D-D reaction. The plasma discharge approach can also be extended to other nuclear reactions such as the p+D reaction.

## 5. Neutron yield results and discussion

In Figure 9, we show relative neutron yields measured during extended plasma discharge runs as a function of cathode voltage. Here, we normalized neutron data for a series of cathode voltages to the data at the highest cathode voltage used in our experiments (i. e. $V_c$= - 12.6 kV). The cathode voltage sets the maximum deuterium ion energy in the laboratory frame, $E_{max}$, after a small correction due to the plasma potential, $E_{max}=q_e$ ($V_p$-$V_c$). The ion energy distribution striking the target is determined through collisions in the sheath that shift the maximum ion energy by an amount dE, as discussed above (Figure 6). Hence, $E_{lab} = E_{max}$ - dE, and $E_{cm}= E_{lab}/2$.

The energy dependent thick target neutron yield, $Y_n(E)$, can be expressed as [1, 8]:

$$Y(E_{max}) = \int_0^{E_{max}} f(E) \int_0^E N_d\, \sigma(\tilde{E}) \left(\frac{d\tilde{E}}{dx}\right)^{-1} d\tilde{E}\, dE \qquad (Eq.\ 2)$$

With deuterium number density, $N_d$, in (d-atoms/cm$^2$), energy dependent fusion reaction cross-section in the neutron channel, $\sigma$ (E), and ion energy loss function dE/dx, from SRIM. Here, we applied a commonly used depth-energy substitution [1, 6, 8, 9]. We also use the ion energy distribution f(E) from the plasma simulations above (Figures 5 and 6). For comparison and to highlight the effect of the ion energy distribution we also compare to yield calculations were we assume that all ions have the maximum ion energy, $E_{max}$. We did not measure the deuterium number density in situ and assume that is the same for all measurements.

For fusion of bare deuterium nuclei the cross section for the neutron channel is 33 micro barn at 6.3 keV in the center of mass frame, or 12.6 keV in the laboratory frame [4]. We compare our data of relative neutron yields to theoretical predictions assuming collisions of bare nuclei [4] and to theory predictions that include a correction to the ion energy from electron screening potentials $U_e$= 400, 750 and 1250 eV. At the lowest ion energies, we observe relative neutron yields that are a factor of 160 to 1000 times higher than would be expected from the bare nuclei D-D fusion cross section, depending on the neutron data analysis method used. The statistical uncertainty in our data was below 10%, even for the runs at the lowest discharge biases and the lowest neutron rates where signal to background ratios were still higher than 2:1. We include yield results following four PSD analysis procedures discussed above for comparison and see that their results differ by a factor of six for the measurement at $V_c$=2.4 keV where neutron yields and neutron/gamma ratios are the lowest. The widely used charge integration method resulted in the highest neutron yields compared to three variants of label-spreading algorithms in our analysis. Raw data and scripts will be made available upon request.

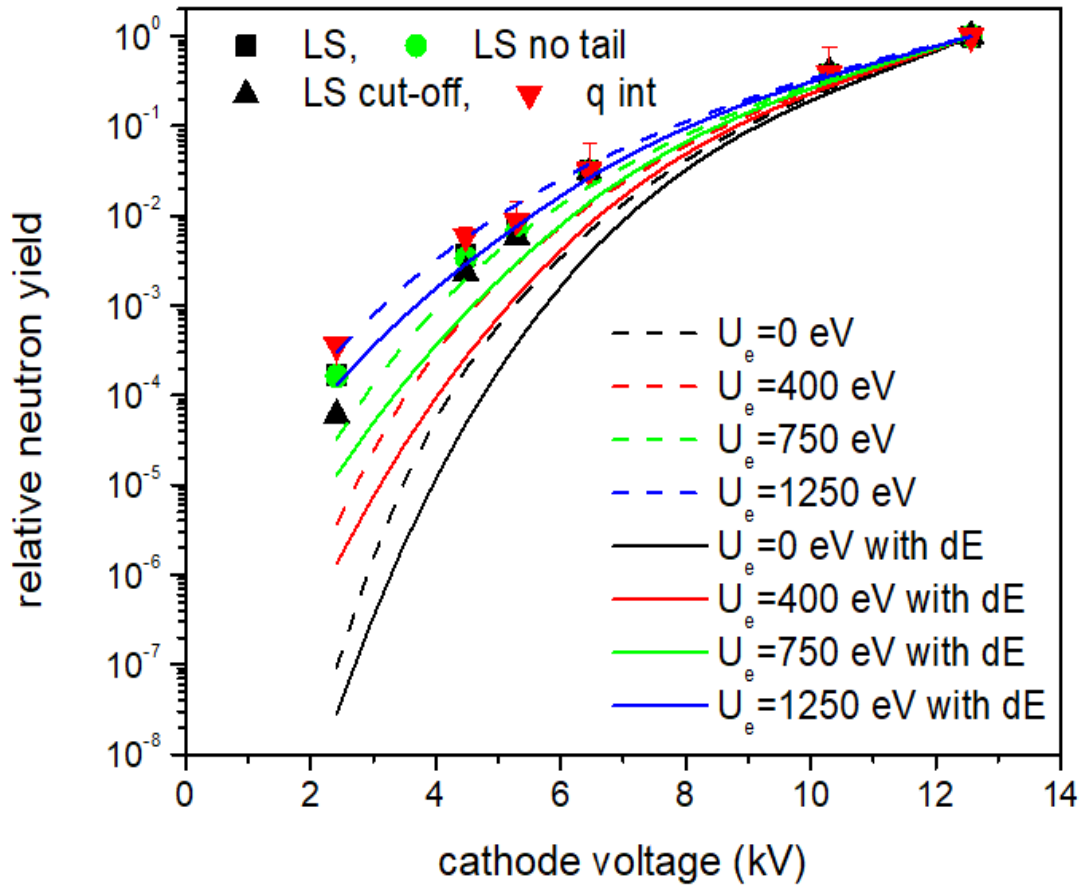

**Figure 9**: Relative thick target yields of detected neutrons as a function of cathode voltage, from four neutron to gamma discrimination procedures discussed in the text (red, down triangles: charge integration; black up triangles: machine learning - label spreading with increased cut-off, black squares: machine learning - label spreading, green dots: machine learning - label spreading with "no tail"). We compared relative yield data to relative yields calculated for bare nuclei [4] (black), and with electron screening potentials $U_e$=400 (red), 750 (green) and 1250 (blue) eV. We include predicted yield curves assuming the calculated ion energy distributions from Figures 6 (solid) and assuming that all ions have the maximum energy given by the cathode voltage (dashed). Systematic errors and biases are discussed in the text and are summarized in Appendix 1. We include an error of a factor of 2 (an upper limit estimate from uncertainties in ion current measurements), in the neutron yields from charge integration.

In our experiments the $D_2$ pressure was 0.5 Torr, except for the lowest energy point ($V_c$=2.4 kV) where it was 2 Torr. The neutron rate at this lowest deuterium ion energy was over 3 neutrons/hour after PSD analysis that gave the lowest neutron counts and the run spanned 70 h. The background rate was 1.6 neutrons/h. Detected neutron count rates were normalized to the integrated discharge current during runs. We did not correct the discharge current for secondary electron emission from the cathode wire; we approximate the discharge current to be the deuterium ion current. The yields of secondary electrons emitted from (clean) metal surfaces from the impact of hydrogen ions with kinetic energies of 2 to 10 keV increase from 0.45 to 1.25 electrons/ion (for molybdenum) [42]. But secondary electron emission is very sensitive to the surface work function (or electron affinity for oxides) and the electronic energy loss of ions (which depends on the target composition) both of which were not measured during our experiments [43]. This could lead to an uncertainty in the ion currents between low and high cathode voltages of up to an estimated factor of two.

The data in Figure 9 were collected with the same palladium wire (0.5 mm diameter) over a period of several weeks. The total accumulated deuterium ion fluence was ~$10^{21}$ cm$^{-2}$. The range of 12.5 keV deuterium ions in palladium (density 12 g/cm$^3$) is about 100 nm (SRIM estimate [17]). Loading to PdD$_x$, x=1, over a target thickness of 100 nm would be achieved in a few minutes of operation with a peak ion current of 1 A/cm$^2$ (corresponding to a peak ion flux of $6 \times 10^{18}$ ions/cm$^2$) and a duty cycle of $10^{-3}$. However, the redistribution of deuterium into the wire bulk and into vacuum during plasma pulses and during extended discharge plasma operation are not known and were not measured in our experiments.

Together with uncertainties in neutron/gamma discrimination at low yields, the main systematic uncertainty in our data results from the unknown secondary electron yields and ion species distribution that affect ion current measurements. Target loading conditions are also not known. The density and depth distribution of deuterium atoms in the target are not known, leading also to uncertainties in ion energy loss and ranges for a series of ion energies. The ion range can be estimated with SRIM simulations and the range of deuterium ions with an incident energy of 5 keV in Pd is about 38 nm while for PdD$_{x=1}$ (where x is the atomic fraction of deuterium nuclei in the Pd matrix) it would be expected to be 70 nm due to the reduced density.

Exact ion energy and species distributions are also unknown. We have used the PIC simulations described above to better understand the plasma conditions but these require experimental benchmarking and validation [34]. For the exponentially increasing D-D cross section for bare nuclei, ions with the highest energies will likely dominate the observed thick target yields. We thus included yield calculations were we assume that all ions are at the maximum energy set by the cathode voltage for comparison in Figure 9. We do not know the exact weight of these uncertainties in the trends of normalized yields. Many of the uncertainties cancel out in relative measurements including measurements of branching ratios. We summarize error contributions in Appendix 1.

With the present spread in the neutron yield data from the analysis methods discussed above relative yield curves for $U_e$ = 750 to 1250 eV show the best agreement to the data over the energy range in our experiments. We estimate $U_e$ = 1000±250 eV for our results in the neutron channel for the D-D reaction, which is consistent with earlier results from measurements in the proton channel [6, 8]. A broad range of values of the electronic screening potential, $U_e$, of ~100 to 800 eV have been reported in D-D fusion experiments with a series of metals, compounds and experimental conditions [5-10] and there have been extensive discussions of the pitfalls and required controls of experimental conditions to achieve reliable conclusions [9]. We argue that $U_e$ is likely dependent on details of target loading and defect dynamics, which are affected and can possibly be controlled in plasma discharges in a regime of high flux (ions/cm$^2$/s) and fluences (ions/cm$^2$) of atomic and molecular ions.

## 6. *Ex situ* analysis of reaction products and cathode wire samples

Following extended plasma exposures of Pd and Ti wires, we extracted the wires and inserted them together with control samples into a liquid scintillation counter (Packard Tri-Carb model B4430, Perkin Elmer) to check for tritium activity through detection of $\beta^-$ (18.59 keV) emission. Due to the expected low tritium activity, samples were counted using the Ultima Gold$^{TM}$ LLT scintillation cocktail ("Low Level of Tritium", Perkin Elmer, Shelton, CT, USA) and with prolonged acquisition time (up to 12 hours). We also did the same for an aluminum catcher target that we had placed into the plasma chamber facing the wire cathode. No activity above background (13 counts per minute) was detected for either sample. With a specific activity of tritium of $3.6 \times 10^{14}$ Bq/g, we estimate an upper bound of tritium in the top layer (with thickness limited to about 300 nm by the escape depth of the 18.59 keV electron from beta-decay of tritium) to be about $10^8$ tritium atoms. *In situ* tracking of tritium activity is clearly preferable, for example based on scintillation counting [16] or mass spectrometry.

Following extended plasma exposures, we have conducted *ex situ* microstructural and compositional examinations of wire cathodes using standard electron microscopy tools. In Figure 10 we show electron micrographs of a Pd wire that had been exposed to total deuterium ion fluence of about $10^{21}$ ion/cm$^2$ during a series of discharge runs. This fluence regime is of interest for fusion reactor development, were high fluences and high fluxes of low energy ions on plasma facing components will be present posing significant materials engineering challenges [44]. The plasma exposed wire shows surface roughness and microstructures as a result of ion exposure and heating during ion pulses. Detection of X-rays induced by high energy electrons (15 keV) in a scanning electron microscope showed a surface composition of palladium with significant contributions from carbon (30 at%) and oxygen (20 at%) in the top 100 nm. The control sample wires had a near-surface composition with smaller contributions from oxygen and carbon (14 at% each). Enhanced oxidation of the Pd wire during extended plasma exposure and ion bombardment can be due to the presence of oxygen from water at the base pressure in the $10^{-7}$ Torr range and potentially enhanced chemisorption and desorption in the presence of the plasma. Surface sputter yields for palladium under the impact of 5 keV deuterium ions are about 0.02 atoms/ion (from SRIM [17], and reference data [45]. A rough estimate of the surface adsorption rate (at room temperature) is 1 monolayer/s, or $10^{15}$ atoms/cm$^2$/s, at a water partial pressure of $10^{-6}$ Torr. During a 20 μs pulse with a deuterium ion current of 0.5 A/cm$^2$ about $10^{12}$ atoms are sputtered off the wire cathode, and at a repetition rate of 50 Hz, $5 \times 10^{13}$ atoms are sputtered per second. Hence the surface sputter rate is likely lower than the rate of re-adsorption. Not including re-adsorption and surface oxidation, a rough estimate of the total material removal from sputtering during extended plasma runs (integrated fluences in the $10^{21}$ ions/cm$^2$ range) is a layer with a thickness of a few μm.

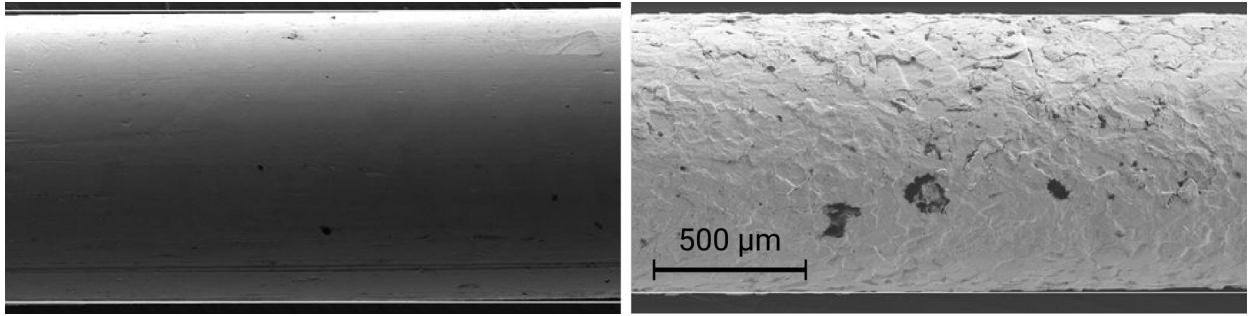

**Figure 10**: Scanning electron micrographs of an as-received Pd wire (left) and a Pd wire that had been exposed to deuterium ions in extended discharge plasma runs (right), imaged using an 15kV, 1 nA electron beam at 57x.

We did not track the wire temperature during plasma runs. With a peak power of 5 kW/cm$^2$ for a current density of 1 A/cm$^2$ and an ion energy of 5 keV the average power at the $10^{-3}$ duty cycle was 5 W. Analytical estimates given the thermal conductivity of a palladium wire indicate a temperature increase of ~100 K during extended operation. Tracking microstructure evolution and surface composition for a series of discharge conditions and looking for possible correlations with fusion reaction rates due to potential changes in electronic screening conditions is the subject of ongoing studies. Here, we can vary the duty cycle or ion current for a given discharge bias and plasma composition to see if fusion rates depend on factors other than the ion energy. But this will also require improved knowledge of target loading conditions, e.g. through implementation of target loading techniques that are independent of beam and plasma conditions together with operando monitoring of deuterium concentrations.

## 6. Conclusions

Plasma discharges enable parametric studies of light ion fusion reactions at relatively low energies and with relatively high peak ion current densities in a compact setup. This enables access to the physics of electron screening effects for varying sample and plasma discharge conditions. We report results for D-D fusion with palladium wires. For nuclear astrophysics where absolute cross sections are required, the main drawbacks of this approach are the uncertainties in target loading conditions, ion species and in the ion energy distributions, and the resulting uncertainty in ion currents. Compared to experiments with well-defined ion beams the ion energy distribution from the plasma discharge is broader and both atomic and molecular ion species can be present simultaneously. Simulations can predict corrections to ion energies compared to applied discharge biases but they have to be benchmarked and validated. Precise tracking of plasma and ion parameters is challenging when plasma discharge conditions are changed. Uncertainties in target loading conditions are common for the plasma and ion beam approaches.

Electron screening effects enhance fusion rates by factors of over 100 at low reaction energies compared to theory predictions for bare nuclei. We report results with cathode voltages from 2.4 to 12.6 kV and corresponding ion energies as low as 1.2 keV (center of mass frame) with conventional neutron detectors and stable operation of the experiment for several days. Extension to lower ion energies is possible with improved detectors. We discuss the use of a series of analysis methods for determination of neutron rates in the presence of gamma ray background. From comparison to yield predictions with a series of values for the electron screening potential we estimate $U_e$=1000±250 eV in our experiments. In a simple model, this correction to the ion kinetic energy increases the probability to tunnel through the repulsive Coulomb barrier. But an electron screening potential of ~1000 eV is not consistent with established theories of electron screening, which reproduce measured values from gas phase experiments of ~27 eV [1, 5-11]. The value of $U_e$ from our measurements in the neutron channel is consistent with earlier results in a similar glow discharge plasma regime and measurements of protons from D-D fusion reactions [8]. We did not detect any tritium in *ex situ* analysis of palladium or titanium cathode wires. A tentative conclusion is that in the energy range probed here the branching ratio between the p+T and n+$^3$He channels does not deviate strongly from the value of approximately one that is well known for D-D fusion reactions [4, 18]. Branching ratios can be determined with future implementation of a proton detector. Plasma discharges offer ways to study and potentially control conditions that can affect electron screening, such as ion dose rates and defect density in the target. With increased understanding of electron screening effects, proposed (sub-)threshold resonances and changes in branching ratios can become accessible using plasma discharges at low reaction energies for a series of nuclear reactions that are relevant for nuclear astrophysics and stellar environments [1, 14, 40, 41]. Compact, high current plasma discharge devices also enable parametric studies of materials relevant for our understanding of plasma-wall interactions in fusion reactors, complementing efforts conducted with large plasma devices [45]. This approach can further inform the development of very compact neutron sources that can replace radiological neutron sources for applications requiring relatively low neutron yields [46].

**Acknowledgments**

Work at Lawrence Berkeley National Laboratory (LBNL) was funded by Google LLC under CRADA (Cooperative Research and Development Agreements, FP00004841, FP00007074 and FP00008139) between LBNL and Google LLC. LBNL operates under U.S. Department of Energy contract DE-AC02-05CH11231.

We thank Takeshi Katayanagi and Peter "Chip" Kozy for technical support. We thank Dave Fork, Ross Koningstein and Matt Trevithick (Google LLC) for stimulating discussions.

Raw data and analysis scripts will be made available upon request.

# Appendix 1

Table 1: Summary of errors and uncertainties affecting measured neutron yields, relative yields and the value of an electron screening potential, $U_e$.

| Parameter | Approach | Comment |
| --- | --- | --- |
| **Ion energy (KeV/u)** | Not measured. Simulations of ion energy distributions with energy loss from collisions in the plasma sheath | The maximum ion energy is given by cathode voltage |
| **Ion species distribution** | Not measured. Simulations of ion species distributions | |
| **Ion current (mA)** | Measured currents are from the sum of ions and secondary electrons | Secondary electron yields depend on ion energies and species distribution and surface conditions, which can vary with plasma conditions. Estimated error is up to a factor of two. |
| **Deuterium concentration (atoms/cm³)** | Not measured | Loss rate into vacuum and diffusion into the bulk of the wire are not known. |
| **Neutron yield** | Neutron gamma pulse shape discrimination with different methods | Spread by a factor 6 for the lowest cathode voltage |
| **Thick target neutron yield** | Estimated yield uncertainty is up to a factor of two due to uncertainty in relative ion currents for different cathode voltages and plasma conditions and a factor of 6 from neutron-gamma discrimination at the lowest cathode voltage (2.4 kV) | Comparison of relative yields to expected yields with maximum ion energy from cathode voltage and comparison to yields from bare nuclei supports the estimate of an electron screening potential $U_e$=1000 +/-250 eV. |

# References

[1] C. E. Rolfs and W. Rodney, Cauldrons of the Cosmos (University of Chicago Press, Chicago, 1988).

[2] D. Casey, et al., Nature Physics 13, 1227 (2017)

[3] Y. Wu and A. Pálffy, The Astrophysical Journal 838:55, 1 (2017)

[4] H. S. Bosch and G. M. Hale, Nuclear Fusion 32, 611 (1992)

[5] K. Czerski, A. Huke, A. Biller, P. Heide, M. Hoeft, G. Ruprecht, Europhys. Lett. 54, 449 (2001); K. Czerski, D. Weissbach, A. Kilic, G. Ruprecht, A. Huke, M. Kaczmarski, N. Targosz-Ślęczka, and K. Maass, Europhys. Lett. 113, 22001 (2016)

[6] F. Raiola, et al., Eur. Phys. J. A 13, 377 (2002)

[7] J. Kasagi, H. Yuki, T. Baba, T. Noda, T. Ohtsuki, A. G. Lipson, J. Phys. Soc. Japan 71, 2881 (2002)

[8] A. G. Lipson, A. S. Rusetski, A. B. Karabut, and G. Miley, Journal of Experimental and Theoretical Physics 100, 1175 (2005).

[9] A. Huke, K. Czerski, P. Heide, G. Ruprecht, N. Targosz, W. Zebrowski, Phys. Rev. C 78, 015803 (2008)

[10] H. J. Assenbaum, K. Langanke, and C. Rolfs, Z. Phys. A - Atomic Nuclei 327, 461 (1987)

[11] S. Ichimaru, Rev. Mod. Phys. 65, 252 (1993)

[12] A. Tumino, et al., Nature 557, 687 (2018).

[13] A Tumino, et al., J. Phys. Conf. Series, 665, 012009 (2016)

[14] M. Wiescher, F. Käppeler, K. Langanke, Annual Review of Astronomy and Astrophysics 50, 165 (2012)

[15] V. Bystritsky, et al., Nucl. Instr. Meth. A 764, 42 (2014).

[16] T. Claytor, D. Jackson, and D. Tuggle, Tritium production from a low voltage deuterium discharge on palladium and other metals, Tech. Rep. (Los Alamos National Lab., NM (United States), 1995).

[17] J. F. Ziegler, M. D. Ziegler and J. P. Biersack, Nucl. Instr. Meth. B 268, 1818 (2010); http://www.srim.org/

[18] P. R. Goncharov, Atomic Data and Nuclear Data Tables 120, 121 (2018)

[19] A. Friedman, et al., IEEE Trans. Plasma Science 42, 1321 (2014); also http://warp.lbl.gov

[20] Tech-X Corporation, https://www.txcorp.com/vsim accessed 29 August 2018

[21] E G Thorsteinsson and J T Gudmundsson, Plasma Sources Sci. Technol. 19, 015001 (2010)

[22] B. G. Lindsay, and R. F. Stebbings, J. of Geophysical Research: Space Physics 110,.A12 (2005)

[23] A. V. Phelps, Journal of Physical and Chemical Reference Data 19, 653 (1990)

[24] J.-S. Yoon, et al., Journal of Physical and Chemical Reference Data, 2008. 37(2): p. 913-931.
Page 21 of 22

[25] R. K. Janev, W. D. Langer, and E. Douglass Jr, "Elementary processes in hydrogen-helium plasmas: cross sections and reaction rate coefficients", Springer Science & Business Media, V 4, 1987

[26] C. Gorse, R. Celiberto, M. Cacciatore, A. Lagana, M. Capitelli, Chemical Physics. 161, 211 (1992)

[27] H. C. Straub, P. Renault, B. G. Lindsay, K. A. Smith, R. F. Stebbings, Physical Review A 54, 2146 (1996)

[28] J. M. Wadehra, J. N. Bardsley, Physical Review Letters. 41, 1795 (1978)

[29] S. F. Biagi, Fortran program, MAGBOLTZ, versions 8.9 and after, (2012) Plasma Data Exchange Project https://fr.lxcat.net/cache/5b980fe8e8834/

[30] H. Tawara H, Y. Itikawa, H. Nishimura, M. Yoshino, J. of Physical and Chemical Reference Data 19, 617 (1990)

[31] P. L. Gertitschke, W. Domcke, Phys. Rev. A 47, 1031 (1993)

[32] H. Gao, Phys. Rev. A 45, 6895 (1992)

[33] L. Marques, J. Jolly, L. L. Alves, J. Appl. Phys. 15, 063305 (2007)

[34] J. Carlsson, A. Khrabrov, I. Kaganovich, T. Sommerer, D. Keating, Plasma Sources Science and Tech. 26, 014003 (2017)

[35] G. F. Knoll, Radiation Detection and Measurement, third ed, Wiley, New York, 2002.

[36] https://eljentechnology.com/products/liquid-scintillators/ej-301-ej-309.

[37] Q. Ji, C. J. Lin, C. Tindall, M. Garcia-Sciveres, T. Schenkel, and B. A. Ludewigt, Rev. Sci. Instr. 88, 056105 (2017).

[38] A.C. Kaplan, M. Flaska, A. Enqvist, J.L. Dolan, S.A. Pozzi, Nucl. Inst. Meth. A 729 (2013) 463

[39] D. Zhou, O. Bousquet, T. Lal, J. Weston, B. Schölkopf, "Learning with Local and Global Consistency," Proceedings of the 16th International Conference on Neural Information Processing Systems, Whistler, BC (2003), http://dl.acm.org/citation.cfm?id=2981345.2981386

[40] M. Kaczmarski, K. Czerski, D. Weissbach, A. I. Kilic, G. Ruprecht, A. Huke, Acta Physica Polonica B 48, 489 (2017)

[41] F. E. Cecil, H. Liu, J. S. Yan, G. M. Hale, Phys. Rev. C 47 1178 (1993)

[42] E. W. Thomas, ed., Particle interactions with surfaces, Vol. 3, Atomic data for Fusion (Oak Ridge National Laboratory, 1985)

[43] Baragiola R.A., Riccardi P. (2008) Electron Emission from Surfaces Induced by Slow Ions and Atoms. In: Depla D., Mahieu S. (eds) Reactive Sputter Deposition. Springer Series in Materials Science, vol 109. Springer, Berlin, Heidelberg

[44] C. E. Kessel, et al., Fusion Engineering and Design 135, 236 (2018)

[45] Y. Yamamura and H. Tawara, Atomic Data and Nuclear Data Tables 62, 149 (1996)

[46] A. Persaud, et al., Rev. Sci. Instr. 83, 02B312 (2012)